\documentclass[twocolumn,aps,superscriptaddress,prb]{revtex4-1}
\usepackage{graphicx}
\usepackage{amssymb}
\usepackage{amsmath}
\usepackage{bm}
\usepackage{color}
\usepackage{epstopdf}

\usepackage[normalem]{ulem} % Strikethrough text (can be removed in the final version)

\def\Vec#1{{\bf #1}}
\def\GVec#1{\mbox{\boldmath $#1$}}

\begin{document}

\title{Quasicrystalline electronic states in 30$^\circ$ rotated twisted bilayer graphene\\
}

\author{Pilkyung Moon}
\thanks{These authors contributed to the manuscript extensively; Corresponding author: pilkyung.moon@nyu.edu}
\affiliation{Arts and Sciences, NYU Shanghai, Shanghai, China; NYU-ECNU Institute of Physics at NYU Shanghai, Shanghai, China}
\affiliation{Department of Physics, New York University, New York, USA}
\affiliation{State Key Laboratory of Precision Spectroscopy, East China Normal University, Shanghai, China}

\author{Mikito Koshino}
\thanks{These authors contributed to the manuscript extensively; Corresponding author: pilkyung.moon@nyu.edu}
\affiliation{Department of Physics, Osaka University, Toyonaka, Japan}

\author{Young-Woo Son}
\affiliation{Korea Institute for Advanced Study, Seoul, Korea}

\begin{abstract}
%Quasicrystals show perfect long-range rotational structural order that is incompatible with the translational crystal symmetry. 
The recently realized bilayer graphene system with a twist angle of $30^\circ$
offers a new type of quasicrystal which unites the dodecagonal quasicrystalline nature and graphene's relativistic properties.
Here, we introduce a concise theoretical framework that fully respects both the dodecagonal rotational symmetry and the massless Dirac nature,
to describe the electronic states of the system.
We find that the electronic spectrum consists of resonant states labeled by 12-fold quantized angular momentum,
together with the extended relativistic states. The resulting quasi-band structure is composed of the nearly flat bands with spiky peaks in the density of states,
where the wave functions exhibit characteristic patterns which fit to the fractal inflations of the quasicrystal tiling.
We also demonstrate that the 12-fold resonant states appear as spatially-localized states in a finite-size geometry,
which is another hallmark of quasicrystal.
The theoretical method introduced here is applicable to a broad class of ``extrinsic quasicrystals"
composed of a pair of two-dimensional crystals overlaid on top of the other with incommensurate configurations.	
\end{abstract}

\maketitle

\section{Introduction}
% introduction: uniqueness, overview and strong points of our arguments

When two graphene layers are overlapped on top of the other, 
the interlayer twist angle $\theta$ is an important physical quantity to determine the electronic structures.
This twisted bilayer graphene (TBG) is essentially a quasi-periodic system,
as the two lattice periods of individual graphene layers are generally irrational to each other.
When $\theta$ is relatively small (less than about 10$^\circ$), however,
the low-energy physics is governed by the long-range moir\'e interference pattern,
and then the electronic properties are captured by the moir\'e effective theory that does not need an exact lattice matching.
In brief, the effective theory approximately treats TBG as a  translationally-symmetric system ruled by the moir\'e period.
%and its effective band structure is defined on the corresponding Brillouin zone.
The exotic phenomena in the low-angle regime \cite{mele2010commensuration,fu2018magic}, such as the flat band formation \cite{lopes2007graphene,trambly2010localization,shallcross2010electronic,morell2010flat,bistritzer2011moire,moon2012energy,de2012numerical}
and the Hofstadter butterfly under magnetic field
\cite{moon2012energy,Dean2013,Hunt2013,Ponomarenko2013,Moon2014}, can be understood in terms of the moir\'e effective theory.

\begin{figure*}
	\centering
	\includegraphics[width=1.\hsize]{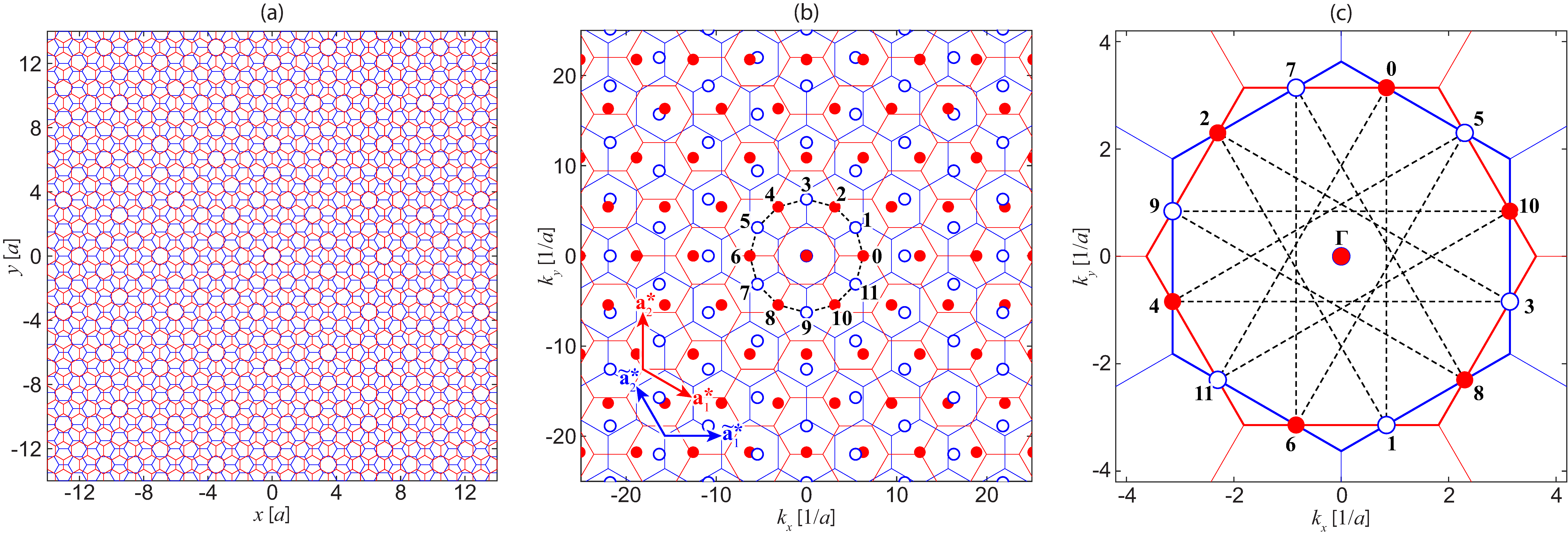}
	\caption{
		(a) Real-space lattice structures of QC-TBG (TBG stacked at 30$^\circ$). Red and blue hexagons represent the graphene's honeycomb lattices of layer 1 and 2, respectively.
		(b)
		Dual tight-binding lattice in the momentum space
		for QT-TBG (see text). 
		Red and blue hexagons show the extended Brillouin zones of layer 1 and 2, respectively.
		The red filled circles represent the wavenumbers $\Vec{k}$ for layer 1,
		and blue open ones represent the inverted wavenumbers $\Vec{k}_0 - \tilde{\Vec{k}}$ for layer 2,
		where $\Vec{k}_0$ is taken as $\Vec{0}$ here.
		The number $n$ represents the position of $\Vec{Q}_n$ $(n=0,1,2,\cdots, 11)$,
		and the dashed line indicates the connection in the 12-ring effective Hamiltonian.
		(c) The original positions of $\Vec{k}$ (layer 1) and $\tilde{\Vec{k}}$ (layer 2) associated with $\Vec{Q}_n$,
		in the first Brillouin zone.
		The dashed line indicates the connection in the 12-ring Hamiltonian as in (a).
		Due to the symmetry, these twelve wavenumbers are at the same distance from the Dirac point
		so that the intrinsic graphene's Bloch states at these wavenumbers are all degenerate in energy.
	}
	\label{fig_lattice}
\end{figure*}

In TBG of large $\theta$, on the other hand, the moir\'e period competes with the atomic length scale
and the quasi-periodic nature emerges \cite{Koren2016}. % the effective approach sketched above breaks down.
When $\theta=30^\circ$, in particular, the overlaid two hexagonal lattices
is mapped onto a 12-fold rotationally symmetric quasicrystalline lattice 
without any translational symmetry [Fig.\,\ref{fig_lattice}(a)],
as first shown by Stampfli \cite{stampfli}. 
Recently, the TBG with a precise rotation angle of $30^\circ$ was experimentally realized and its spectrum measured in epitaxially grown samples on top of SiC surface \cite{ahn2018dirac}. In addition, similar TBGs have been realized on top of Ni surface \cite{Takesaki2016, Yao2018} and also by a transfer method \cite{Chen2016}. Moreover, another $30^\circ$-rotated stack of atomic layers have also been realized in graphene on top of BN layer \cite{Wang2016} as well as MoSe$_2$ bilayer system \cite{Choi2017}.
In such the quasicrystalline TBG (QC-TBG), the moir\'e effective approach sketched above breaks down
because its main assumption that the moir\'e pattern governs the system
is no longer valid.
%Therefore, the quasicrystalline TBG (QC-TBG) that is the TBG with $\theta=\pi/6$ violates 
%the main assumption of the effective theory regarding on the symmetry of effective BZ where the theory works.
%Since the hexagonal symmetry of moir\'e lattice breaks down here, a new effective 
%theory should encompass the 12-fold rotational lattice symmetry without translational symmetry.

In the literature, several theoretical approaches have been applied to understand the electronic structures of conventional quasi-periodic systems \cite{Roche1997},
such as one-dimensional Fibonacci lattices \cite{Niu1986,kohmoto1987critical}, two-dimensional non-periodic tiling including Penrose lattice \cite{niizeki1990reciprocal,gambaudo2014brillouin}, metal nanoparticles \cite{Dong2009}, photonic quasicrystals \cite{Mnaymneh2007} and 
three-dimensional alloys including Al-Mn, Al-Ni-Co, and Al-Cu-Co \cite{smith1987pseudopotentials,Fujiwara1991,Hafner1992,TramblydeLaissardiere1994,Roche1998,rogalev2015fermi}.
These systems can be viewed as \textit{intrinsic} quasicrystals where the atomic sites are intrinsically arranged in the quasi-periodic order.
In contrast, the QC-TBG is regarded as an \textit{extrinsic} quasicrystal, in that it is composed of a pair of perfect crystals having independent periodicities,
and the quasi-periodic nature appears only in the perturbational coupling between the two subsystems.
Thus, the QC-TBG unites the quasicrystalline order and the relativistic nature of the massless Dirac particles of graphene,
yet it is not obvious whether and in what form the essential features of quasicrystals emerge in the electronic properties.
Since such a hybrid situation is out of the scope of the previous theories of intrinsic quasicrystals,
we need an alternative theoretical framework to properly describe the quasicrystalline physics of QC-TBG.

%One of the earliest tries is based on a perturbative treatment of nearly free electron states with respect to pseudopotentials reflecting
%geometrical factors with higher dimensional reciprocal vectors of quasicrystal. 
%This approach assumes that quantum states of electron in three dimensional metallic quasicrsytals may experience strong perturbations
%to a small fraction of the Fermi volume of metallic states.
%\textcolor{blue}{[Ref: PRB 81, 161405(R), PNAS 108, 12233, J. Phys.: Condens. Matter 2, 2759, PRL 59, 1365, Nat. Comm. 6, 8607, NJP 16, 043013]} 
% The other popular method is based on rational approximants of the highest possible quasicrystal-like local order within limitation of computational power. 
% This has some merits when treating realistic atomic compositions. However, the essential quasiperiodic nature is absent because of imposed artificial periodicities.
%\textcolor{red}{(TODO: Mention finite flakes.)}
 %The both considerations have shown some characteristics of quasicrystals but, by and large, they are neither concise nor 
 %comprehensive theoretical tools capturing all essential features such as coexistence of critical, localized, extended electronic states and 
 %other anomalous experiment observations
%shown in quasicrystals.
% \textcolor{red}{(Note: Currently, there is no physically meaningful phenomena which cannot be described by the approximant. (Well, finite flake cannot give the band dispersion.))}

In this paper, we develop a concise model Hamiltonian that fully respects both the dodecagonal rotational symmetry and the massless Dirac nature, 
to describe the quasicrystalline electronic states in the QC-TBG. 
We find that the electronic spectrum of QC-TBG is characterized by the 12-fold resonant states of relativistic Dirac fermions,
and they can be well captured by a ring Hamiltonian composed of 12 Dirac cones. 
The resulting quasi-band structure comprises a series of the nearly flat bands corresponding to the resonant states,
each of which is labeled by a 12-fold quantized angular momentum.
The spatial pattern of wave functions exhibit the fractal inflations of the Stampfli tiling,
which is a direct manifestation of the quasicrystalline nature \cite{Niizeki1989}.
Since we can tune the twist angles in the model, 
the transition of electronic states from the approximants \cite{Goldman1993} of QC-TBG to a true dodecagonal rotational symmetry can be
continuously described within a 12-fold ring model,
and the emergence of quasicrystalline states and the validity of the approximant method are critically attested. 
We also show that the 12-fold resonant states appear as spatially-localized states in a finite-size geometry,
which is another hallmark of the quasicrystalline nature \cite{Niu1986,kohmoto1987critical,Deguchi2012}.
The proposed theoretical approach is applicable to a broad class of extrinsic quasicrystals,
and its simple structure of the closed Hamiltonian allows rigorous analysis on exotic quantum phenomena of quasicrystals.
% As the 12-ring Hamiltonian has a closed form unlike any other previous model studies on quasicrystal,
%our model will be able to provide a rigorous analysis on various exotic quantum phenomena of quasicrystals.

The paper is organized as follows. In Sec.\,\ref{sec_theor}, we present the tight-binding model for QC-TBG,
and introduce the dual tight-binding approach in the momentum space.
In Sec.\,\ref{sec_12-fold}, we derive the approximate 12-wave ring Hamiltonian, and using this, we
describe the quasi-band structure, the resonant states and the characteristic wave functions to respect the Stampfli tiling.
In Sec.\,\ref{sec_finite}, we calculate the electronic states of QC-TBG in an alternative method using the finite-size tight-binding model,
and demonstrate the localization nature of the 12-fold resonant states.
A brief conclusion is given in Sec.\ \ref{sec_concl}.

\section{Theoretical methods}
\label{sec_theor}
% atomic structure
\subsection{Tight-binding Hamiltonian for QC-TBG}

We define the atomic structure of QC-TBG 
by starting from AA-stacked bilayer graphene (i.e. perfectly overlapping honeycomb lattices)
and rotating the layer 2 around the center of hexagon by $30^\circ$ [Fig.\,\ref{fig_lattice}(a)].
We set $xy$ coordinates parallel to the graphene layers and $z$ axis perpendicular to
%it
the plane.
The system belongs to the symmetry group $D_{6d}$,
and it is invariant under an improper rotation $R(\pi/6) M_z$, 
where $R(\theta)$ is the rotation by an angle $\theta$ around $z$ axis,
and $M_z$ is the mirror reflection with respect to $xy$ plane. 
The primitive lattice vectors of layer 1 are taken as  $\Vec{a}_1 = a(1,0)$ and $\Vec{a}_2 = a(1/2,\sqrt{3}/2)$
with the lattice constant $a \approx 0.246\,\mathrm{nm}$,
and those of the layer 2 as $\tilde{\Vec{a}}_i = R(\pi/6)\,\Vec{a}_i$.
Accordingly, the reciprocal lattice vectors
of layer 1 are given by  $\Vec{a}^*_1 = (2\pi/a)(1,-1/\sqrt{3})$ and $\Vec{a}^*_2=(2\pi/a)(0,2/\sqrt{3})$,
and layer 2 by $\tilde{\Vec{a}}^*_i = R(\pi/6)\, \Vec{a}^*_i$.
The atomic positions are given by
\begin{align}
&\Vec{R}_{X}=n_1\Vec{a}_{1}+n_2\Vec{a}_{2}+\GVec{\tau}_X
&(\mbox{layer 1}), \nonumber\\
&\Vec{R}_{\tilde{X}}=\tilde{n}_1\tilde{\Vec{a}}_{1}+\tilde{n}_2\tilde{\Vec{a}}_{2}+\GVec{\tau}_{\tilde{X}}
&(\mbox{layer 2}),
\end{align}
where $n_i$ and $\tilde{n}_i$ are integers, 
$X=A,B$ ($\tilde{X}=\tilde{A},\tilde{B}$) denotes the sublattice site of layer 1(2),
and $\GVec{\tau}_X$ and $\GVec{\tau}_{\tilde{X}}$ are the sublattice positions in the unit cell,
defined by $\GVec{\tau}_A=-\GVec{\tau}_1$, $\GVec{\tau}_B=\GVec{\tau}_1$,
$\GVec{\tau}_{\tilde{A}}= - R(\pi/6)\GVec{\tau}_1 + d\Vec{e}_z$, $\GVec{\tau}_{\tilde{B}}=R(\pi/6)\GVec{\tau}_1+ d\Vec{e}_z$
with $\GVec{\tau}_1 = (0,a/\sqrt{3})$.
Here $d \approx 0.335\,\mathrm{nm}$ is the interlayer spacing between graphene layers and
$\Vec{e}_z$ is the unit vector normal to the layer.

% tight-binding

We model graphene by the tight-binding model of carbon $p_z$ orbitals.
The Hamiltonian is spanned by the Bloch bases of $p_z$ orbitals at difference sublattices,
\begin{align}
	& |\Vec{k},X\rangle = 
	\frac{1}{\sqrt{N}}\sum_{\Vec{R}_{X}} e^{i\Vec{k}\cdot\Vec{R}_{X}}
	|\Vec{R}_{X} \rangle\quad (\mbox{layer 1}), \nonumber\\
	& |\tilde{\Vec{k}},\tilde{X}\rangle = 
	\frac{1}{\sqrt{N}}\sum_{\Vec{R}_{\tilde{X}}} e^{i\tilde{\Vec{k}}\cdot\Vec{R}_{\tilde{X}}}
	|\Vec{R}_{\tilde{X}}\rangle \quad (\mbox{layer 2}),
	\label{eq_bloch_base}
\end{align}
where $|\Vec{R}_{X} \rangle$ is the atomic $p_z$ orbital at the site $\Vec{R}_{X}$, 
$\Vec{k}$ and $\tilde{\Vec{k}}$ are the two-dimensional Bloch wave vectors
and $N = S/S_{\rm tot}$  is the number of graphene's unit cells $S=(\sqrt{3}/2)a^2$
in the total system area $S_{\rm tot}$. 
We assume that the transfer integral between any two $p_z$ orbitals
is expressed as \cite{slater_koster}
\begin{eqnarray}
&& -T(\Vec{R}) = 
V_{pp\pi}\left[1-\left(\frac{\Vec{R}\cdot\Vec{e}_z}{R}\right)^2\right]
+ V_{pp\sigma}\left(\frac{\Vec{R}\cdot\Vec{e}_z}{R}\right)^2,
\nonumber \\
&& V_{pp\pi} =  V_{pp\pi}^0 e^{- (R-a/\sqrt{3})/r_0},
\quad V_{pp\sigma} =  V_{pp\sigma}^0  e^{- (R-d)/r_0},
\label{eq_slater_koster}
\end{eqnarray}
where
$\Vec{R}$ is the relative vector between two atoms,
$V_{pp\pi}^0 \approx -2.7\,\mathrm{eV}$,
$V_{pp\sigma}^0 \approx 0.48\,\mathrm{eV}$,
and $r_0 \approx 0.0453\,\mathrm{nm}$ \cite{trambly2010localization,moon2013optical}.

% Hamiltonian matrix
 
The total tight-binding Hamiltonian is expressed as
$H = H_{1} + H_{2} + U$
where $H_{1}$ and $H_{2}$ are the Hamiltonian for the intrinsic monolayer graphenes of layer 1 and 2, respectively,
and $U$ is for the interlayer coupling.
The intralayer matrix elements of layer 1 are given by
\begin{align}
&\langle\Vec{k}',X' | H_1 |\Vec{k},X\rangle
=  h_{X,X'}(\Vec{k}) \delta_{\Vec{k}', \Vec{k}},
\nonumber\\
&  h_{X,X'}(\Vec{k})  = \sum_{\Vec{L}}
-T(\Vec{L}+\GVec{\tau}_{X'X}) e^{-i\Vec{k}\cdot(\Vec{L}+\GVec{\tau}_{X'X})},
\label{eq_H0}
\end{align}
where $\Vec{L} = n_1 \Vec{a}_1 + n_2 \Vec{a}_2$ and $\GVec{\tau}_{X'X} = \GVec{\tau}_{X'}- \GVec{\tau}_{X}$.
Similarly, the matrix for $H_2$ is given by replacing $\Vec{k}$ with $R(-\pi/6)\Vec{k}$.

The interlayer matrix element between layer 1 and 2
%The interlayer Hamiltonian
%\textcolor{red}{between the Bloch states of incommensurately stacked atomic layers}
is written as \cite{mele2010commensuration,bistritzer2011moire,koshino2015interlayer}
\begin{align}
%U_{\tilde{X}X}(\tilde{\Vec{k}},\Vec{k}) 
%&\equiv
\langle\tilde{\Vec{k}},\tilde{X}| U	|\Vec{k},X\rangle
=
-\sum_{\Vec{G},\tilde{\Vec{G}}}
{t}(\Vec{k}+\Vec{G})
e^{-i\Vec{G}\cdot\mbox{\boldmath \scriptsize $\tau$}_{X}
	+i\tilde{\Vec{G}}\cdot\mbox{\boldmath \scriptsize $\tau$}_{\tilde{X}}}
\, \delta_{\Vec{k}+\Vec{G},\tilde{\Vec{k}}+\tilde{\Vec{G}}},
\label{eq_matrix_element_of_U}
\end{align}
where $\Vec{G}=m_1 \Vec{a}^*_1 + m_2 \Vec{a}^*_2$ and
$\tilde{\Vec{G}}=\tilde{m}_1 \tilde{\Vec{a}}^*_1 + \tilde{m}_2 \tilde{\Vec{a}}^*_2$
($m_1, m_2, \tilde{m}_1, \tilde{m}_2 \in \mathbb{Z}$)
run over all the reciprocal points
of layer 1 and 2, respectively.  We also defined
\begin{eqnarray}
{t}(\Vec{q}) = 
\frac{1}{S} \int
T(\Vec{r}+ z_{\tilde{X}X}\Vec{e}_z) 
e^{-i \Vec{q}\cdot \Vec{r}} d\Vec{r}
\label{eq_ft}
\end{eqnarray}
where $z_{\tilde{X}X} = (\GVec{\tau}_{\tilde{X}}-\GVec{\tau}_{X})\cdot\Vec{e}_z$.

%Dual tight-binding model
\subsection{Dual tight-binding lattice in momentum space}
\label{sec_dual_tb}

Equation (\ref{eq_matrix_element_of_U}) shows that the interlayer interaction occurs
between the states satisfying
the generalized Umklapp scattering condition $\Vec{k} + \Vec{G} = \tilde{\Vec{k}} + \tilde{\Vec{G}}$.
%\textcolor{blue}{[Ref: PRB 81, 161405(R), PNAS 108, 12233, J. Phys.: Condens. Matter 2, 2759, PRL 59, 1365, Nat. Comm. 6, 8607, NJP 16, 043013]} 
%\begin{equation}
%\Vec{k} + \Vec{G} = \tilde{\Vec{k}} + \tilde{\Vec{G}}.
%\label{eq_generalized_umklapp_scattering}
%\end{equation}
When we start from the layer 1's Bloch states at $\Vec{k}_0$, for example, 
the interlayer Hamiltonian $U$ couples this state with layer 2's Bloch states at
$\tilde{\Vec{k}} = \Vec{k}_0 + \Vec{G} - \tilde{\Vec{G}}$. They are further coupled back to layer 1's states at
$\Vec{k} = \Vec{k}_0 + \Vec{G}' - \tilde{\Vec{G}}'$, and so forth.
As a result,  the space of the wave functions associated with $\Vec{k}_0$ is spanned by
$\{|\Vec{k},X\rangle \, | \, \Vec{k} = \Vec{k}_0 + \tilde{\Vec{G}} - \Vec{G}\}$
and
$\{|\tilde{\Vec{k}},\tilde{X}\rangle \, | \, \tilde{\Vec{k}} = \Vec{k}_0 + \Vec{G} - \tilde{\Vec{G}}\}$
for $\forall \Vec{G}$ and $\forall \tilde{\Vec{G}}$.
However, we actually need only a subset of these sets, since the BZ of each layer is
translationally invariant with respect to the reciprocal lattice vectors of its own
(i.e., $\Vec{k}$ and $\Vec{k} + \Vec{G}$ stand for the same Bloch wavenumber of layer 1).
Thus, without loss of generality, we can choose 
the subspace spanned by the QC-TBG Hamiltonian as
$\{|\Vec{k},X\rangle \, | \, \Vec{k} = \Vec{k}_0 + \tilde{\Vec{G}}, \forall\tilde{\Vec{G}} \}$
and
$\{|\tilde{\Vec{k}},\tilde{X}\rangle \, | \, \tilde{\Vec{k}} = \Vec{k}_0 + \Vec{G}, \forall\Vec{G} \}$.
Here note that
the $k$-points in each layer
is regularly spaced with the reciprocal vectors of the other layer.
% (Figs.\,\ref{fig_umklapp_scattering_in_qc}b and \ref{fig_umklapp_scattering_in_qc}c).

According to Eq.\ (\ref{eq_matrix_element_of_U}), the interaction strength between 
$\Vec{k}=\Vec{k}_0 + \tilde{\Vec{G}}$ and $\tilde{\Vec{k}}= \Vec{k}_0 + \Vec{G}$ is given by $t(\Vec{q})$ where 
$\Vec{q}=\Vec{k}+\Vec{G}= \tilde{\Vec{k}} + \tilde{\Vec{G}}=\Vec{k}+\tilde{\Vec{k}}-\Vec{k}_0$.
Since $t(\Vec{q})$ decays in large $\Vec{q}$,
the relevant contribution occurs only when $|\Vec{k}+\tilde{\Vec{k}}-\Vec{k}_0|$ is relatively small.
The interaction strength can be visualized by the diagram Fig.\,\ref{fig_lattice}(b), where
all the layer 2's wave points $\tilde{\Vec{k}}$ are inverted to 
$\Vec{k}_0 - \tilde{\Vec{k}}$, and overlapped with the  layer 1's wave points $\Vec{k}$.
In the map, the quantity $|\Vec{k}+\tilde{\Vec{k}}-\Vec{k}_0|$ is the geometrical distance by given two points, 
so that the interaction takes place only between the points located in close distance.
If the $k$-points are viewed as `sites', the whole system can be recognized as 
a tight-binding lattice in $k$-space, which is the dual counterpart of the original tight-binding Hamiltonian in the real space.
It should be noted that, unlike the real-space version, 
the intralayer Hamiltonians $H_1$ and $H_2$ now
can be interpreted as $k$-dependent on-site potential in the $k$-space,
which is nothing but the band energy of intrinsic graphene.
Recently, the relationship between the real space and the momentum space was also noticed in the localized wave functions
in moir\'e bilayer systems \cite{carr2018duality}.

In this $k$-space tight-binding model,
the hopping between different $k$-space sites (the interlayer interaction $U$)
is smaller by an order of magnitude than the potential landscape (the band energy),
so that the eigen functions tend to be localized in the $k$-space lattice,
in a similar manner to the Aubry-Andr\'{e} model in one dimension \cite{aubry1980analyticity}.
In the practical calculation, therefore, we are allowed to take only a limited number of wave points around $\Vec{k}_0$ inside a certain cut-off circle,
and obtain the energy eigenvalues by diagonalizing the Hamiltonian matrix within the finite bases.
If we plot the energy levels against $\Vec{k}_0$, we obtain the quasi-band structures of the system.
Here the wavenumber $\Vec{k}_0$ works like the crystal momentum for the periodic system,
so it can be called the quasicrystal momentum.
The cut-off radius $k_c$ should be greater than the typical localization length in the $k$-space,
but need not be too large, since the wave points discarded outside $k_c$
are properly considered by shifting $\Vec{k}_0$.
If we increase $k_c$, we will see more and more replicas of the identical quasi-energy band with different origins,
because shifting $\Vec{k}_0$ actually corresponds to taking a different origin in the $k$-space map of Fig.\,\ref{fig_lattice}(b).
%In the appendix \ref{SI_quasicrystal_approximant}, we present the extensive study on the electronic structures of the quasi approximants
%in all the angle region from $0^\circ$ to $30^\circ$.
%\label{SI_localization_in_momentum_space}
%\label{SI_effects_of_shifting_k0}
The resonant band structure near $\Vec{k}_0=0$ barely changes in this process because its wave function is very well localized to the 12-membered ring in the $k$-space.
The replica bands are just duplication of the same states 
so they should be appropriately removed in calculating the physical quantities such as the density of states.
The validity of the momentum cut-off is discussed in detail in Appendix \ref{SI_validity}.

\begin{figure*}
	\begin{center}
		\leavevmode\includegraphics[width=0.9\hsize]{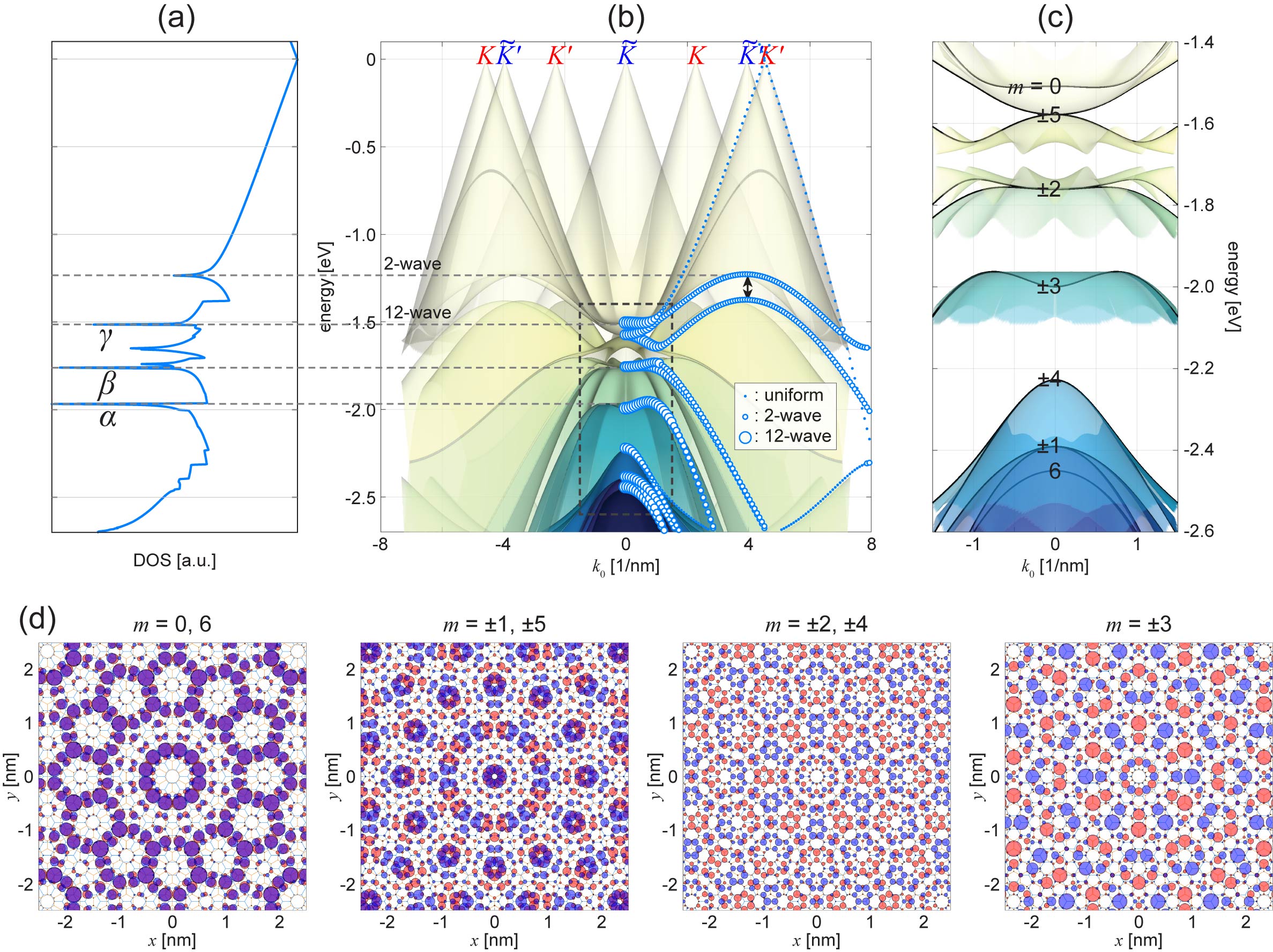}
	\end{center}
\caption{
(a) DOS and (b) electronic structures
in the valence band side of QC-TBG calculated by the 12-ring effective model.
Blue dots represent the inverses participation ratio of the dominant layer
at several sample points in the band structures, where the dot area is proportional to
the measure of the spatial extent of the wave functions.
Inset shows the size of the dots for almost decoupled states (``uniform"),
the states arise from the hybridization of the two-waves in the same layer (``2-wave"),
and the states arise from the hybridization of twelve-waves (``12-wave") (see text).
The black arrow shows the gap opening caused by 2-wave mixing.
%		\textcolor{red}{[Note: \Vec{b} shows only $210^\circ$, 7 Dirac cones,
%			to make the figure clearer, while there are 12 cones in total.]}
(c) Detailed band structures near $\Vec{k}=\Vec{0}$
with index $m$ indicating quantized angular momentum in 12-fold rotational symmetry.
(d) The valence band wave functions at $\Vec{k}=\Vec{0}$ characterized by
$m$, where the area of the circle is proportional to the squared wave amplitude, and red and blue
circles represent the states in the upper and the lower layers, respectively.
}
\label{fig_12g_vb}
\end{figure*}

%%%
\section{Results and discussion}
\subsection{12-fold symmetric resonant states}
\label{sec_12-fold}

At $\Vec{k}_0 = \Vec{0}$, we see that the twelve symmetric points 
$\Vec{Q}_n = a^* [\cos (n\pi/6), \sin (n\pi /6)] (n=0,1,2,\cdots, 11)$
form a circular chain in the dual-tight-binding lattice of which 
radius is $a^*\equiv |\Vec{a}^*_i|=4\pi/(\sqrt{3}a)$, 
indicated by the dashed ring in Fig.\,\ref{fig_lattice}(b).
Noting that the layer 2's wave points are inverted, 
these points are associated with layer 1's Bloch wavenumbers $\Vec{k} = \Vec{Q}_n$ for even $n$'s
and layer 2's $\tilde{\Vec{k}} = -\Vec{Q}_n$ for odd $n$'s.
Figure \ref{fig_lattice}(c) shows 
the original positions of $\Vec{k}$ (layer 1) and $\tilde{\Vec{k}}$ (layer 2) associated with $\Vec{Q}_n$,
 in the first Brillouin zone.
Due to the symmetry, the intrinsic graphene's Bloch states at the twelve points are all degenerate in energy,
and therefore the interlayer coupling hybridizes them to make resonant states.
Here the coupling is only relevant between the neighboring sites of the ring,
and it is given by $t_0 = t(2a^*\sin 15^\circ) \approx 157\,\mathrm{meV}$.

In the vicinity of $\Vec{k}_0=\Vec{0}$, the Hamiltonian of the ring can be expressed by a $24\times 24$ matrix,
\begin{align}
&{\cal H}_{\rm ring}(\Vec{k}_0) = 
\begin{pmatrix}
H^{(0)} & W^\dagger &&&& W \\
W & H^{(1)} & W^\dagger \\
& W & H^{(2)} & W^\dagger \\
&& \ddots & \ddots &\ddots \\
&&& W & H^{(10)} & W^\dagger \\
W^\dagger &&&& W & H^{(11)}
\end{pmatrix},
\label{eq_H12}
\\
&  H^{(n)}(\Vec{k}_0) = 
\begin{pmatrix}
h_{AA}^{(n)} & h_{AB}^{(n)}  \\
h_{BA}^{(n)} & h_{BB}^{(n)}  \\
\end{pmatrix},
%& \qquad  H^{(n)}(\Vec{k}_0) =  H_0[R(-7n\pi/6) \Vec{k}_0 + \Vec{Q}_0]
\quad
W =  - t_0
\begin{pmatrix}
\omega & 1\\
1 & \omega^*
\end{pmatrix},
\end{align}
where $h_{X'X}^{(n)}(\Vec{k}_0) = h_{X'X}[R(-7n\pi/6) \Vec{k}_0 + \Vec{Q}_0]$, $\omega=e^{2\pi i/3}$,
and we neglect the $\Vec{k}_0$ dependence of the interlayer matrix element $t(\Vec{q})$.
The diagonal block $H^{(n)}$ represent monolayer's Hamiltonian at $\Vec{k} = \Vec{k}_0 + \Vec{Q}_n$ for even $n$ (layer 1)
and $\tilde{\Vec{k}} =\Vec{k}_0 - \Vec{Q}_n$ for odd $n$ (layer 2).
In each  $2\times 2$ block the sublattices are arranged in the order of $(A,B)$ or $(\tilde{A}, \tilde{B})$ for $n\equiv 0,3$ in modulo of 4,
and $(B,A)$ or $(\tilde{B}, \tilde{A})$ for $n\equiv 1,2$.
By doing this, the first base of a $2\times 2$ block is always mapped to the first base of other block
under the operation of $R(\pi/6) M_z$.
Note that the arrangement of $h_{AA}$, $h_{AB}$, etc.\ in the submatrix $H^{(n)}$ is fixed irrespective of $n$,
and the dependence on $n$ solely comes from $R(-7n\pi/6) \Vec{k}_0$ in the argument of $h_{X'X}$.
Consequently, the total Hamiltonian ${\cal H}_{\rm ring}$ is obviously symmetric under
rotation by a single span of the ring (i.e., moving $\Vec{Q}_n$ to $\Vec{Q}_{n+1}$),
which actually corresponds to the operation $[R(\pi/6) M_z]^7$ (210$^\circ$ rotation and swapping layer 1 and 2) in the original system.

Figures\,\ref{fig_12g_vb} show (a) the density of states (DOS),
(b) the band structures as a function of $\Vec{k}_\Vec{0}$ in the negative energy region,
and (c) its closer view near $\Vec{k}_0=\Vec{0}$.
The twelve Dirac cones are arranged on a circle with a radius $\Delta k = 4(2-\sqrt{3})\pi/(3a)$
and they are strongly mixed near $\Vec{k}_0=\Vec{0}$.
As a result,
the originally degenerate twelve states of graphene (in each of the electron side and the hole side of the Dirac cone) 
split into different energies, and exhibit the characteristic dispersion
including flat band-bottoms and the Mexican-hat edges.
This leads to a series of spiky peaks and dips (pseudogaps) in DOS.	
At $\Vec{k}_0=\Vec{0}$, the Hamiltonian can be analytically diagonalized to obtain a set of energies (neglecting the constant energy)
\begin{align}
& E^\pm_{m} =  t_0 \cos q_m \pm \sqrt{3t_0^2 \sin^2 q_m + (h_0 - 2 t_0 \cos q_m)^2},
\label{eq_Em}
\end{align}
where $h_0 = h_{AB}(\Vec{Q}_0) = h_{BA}(\Vec{Q}_0) = $ 1.84 eV, $\pm$ corresponds to the conduction band and valence band,
respectively, and $q_m = (7\pi/6) m$ with $m = -5,-4,\cdots, 5,6$ is the wavenumber along the chain.
The eigenvalue of $R(\pi/6) M_z$ is given by $e^{i m\pi/6}$.
Here the states with $m=\pm s$ $(s=1,2,3,4,5)$ form twofold doublets, and belong to
two-dimensional $E_s$ irreducible representation of $D_{6d}$ point group.
The $m=0$ and $6$ are non-degenerate, and belong to $A_1(A_2)$ and $B_2 (B_1)$,
respectively, for the conduction (valence) band.
If we disregard the $z$-position difference,
the index $m$ is regarded as quantized angular momentum in 12-fold rotational symmetry,
and this is an essential characteristics of quasicrystal TBG.

We have similar resonant states also in the conduction band,
while the energy scale of the band structures is much smaller than in the valence band.
	Equation (\ref{eq_Em}) clearly explains such asymmetry; the dispersion of $E^-_{m}$ in $q_m$ is nearly three times wider than that of $E^+_{m}$, considering that $h_0 \gg t_0$.
Intuitively, the wave function of the conduction band of intrinsic graphene
has the opposite phases between the sublattice $A$ and $B$, and then the interlayer coupling between incommensurate layers  
is tend to be suppressed by the phase cancellation.

\begin{figure}
	\begin{center}
		\leavevmode\includegraphics[width=0.75\hsize]{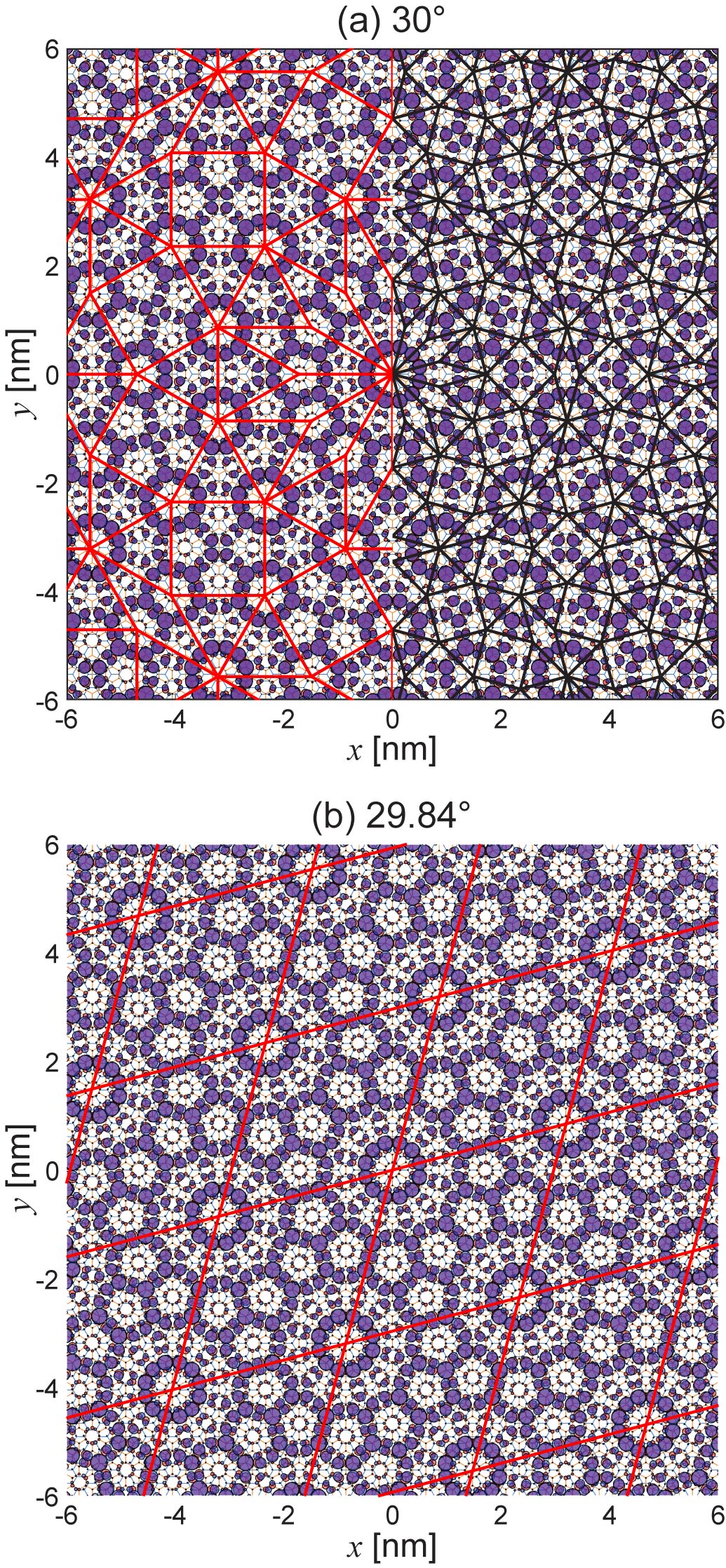}
	\end{center}
	\caption{(a) Large scale plot of $m=0, 6$ states of TBG with $\theta=30^\circ$
		in Fig.\,\ref{fig_12g_vb}(d). Red (left-half) and black (right-half) lines represent 
		the fourth and third generations of the Stampfli tiling, respectively. 
		(b) Similar plot for the quasicrystal approximant with $\theta=29.84^\circ$.
		The red lines indicate the periodic unit cell.
	}
	\label{fig_ldos_symmetry_approximant_and_ecm}
\end{figure}

% real space wave function and IPR
\subsection{Wave functions showing the quasicrystal tiling}

The 12-wave resonant coupling also gives rise to a characteristic pattern in the wave function.
Figure\,\ref{fig_12g_vb}(d) shows the wave functions at $\Vec{k}_0=\Vec{0}$ where the hybridization is the most prominent,
where we can see that the wave amplitude distribute selectively on a limited number of sites in a 12-fold rotationally symmetric pattern.
The extent of the hybridization of different wave modes is characterized by 
the inverse participation ratio (IPR) on the dominant layer, 
\begin{equation}
 P^{-1}(\psi) = \frac{\sum'_i |\psi_i|^4}{\left(\sum'_i |\psi_i|^2\right)^2}.
\end{equation}
Here $\psi_i$ is the amplitude at the site $i$ of the eigenstates $\psi$,
and $\sum'_i$ represents the sum over the sites on the dominant layer,
which is defined as the layer having greater wave amplitude than the other.
We have $P^{-1} = 1$ for a pure single layer state, and  $P^{-1} = 1.5$ for a hybrid state of two plain wave modes.
In Fig.\,\ref{fig_12g_vb}(b), the blue dots represent IPR at several sample points in the band structures along $x$ direction,
where the dot area is proportional to  $P^{-1} -1$.
We see that the IPR becomes large exclusively around $\Vec{k}_0=\Vec{0}$, where the 12 wave components are strongly hybridized.
$P^{-1}$ remains almost 1 near the Dirac cones where the hybridization is almost negligible.
We also have a region of $P^{-1} \sim 1.5$ along the arch-shaped gap below the Dirac cone,
which is indicated by ``2-wave" in Fig.\,3(b).
These ``2-wave" states arise from the hybridization of the $K$ and $K'$ of the {\it same} layer
assisted by the second-order process of the interlayer coupling $U$.

We also show  a large scale plot of $m=0, 6$
states in Fig.\,\ref{fig_ldos_symmetry_approximant_and_ecm}(a).
We see that the wave pattern perfectly follows the Stampfli tiling,
where the red (left-half) and black (right-half) lines represent the third and fourth generations of the fractal inflation, respectively \cite{stampfli}.
Such the long-range structure of the quasicrystalline wave function is actually quite sensitive to a slight change of the twist angle.
%Our momentum-space tight-binding approach works in any twist angles,
%and we can derive the 12-ring effective Hamiltonian as a good approximation for other $\theta$'s near 30$^\circ$.
Figure\,\ref{fig_ldos_symmetry_approximant_and_ecm}(b)
represents the wave pattern of the corresponding state in TBG with $\theta = 29.84^\circ$,
calculated by the same 12-wave method.
The TBG of $29.84^\circ$ is a quasicrystal approximant, which is not quasi-periodic but has a translational symmetry with period of $3.31\,\mathrm{nm}$.
We can see that the local wave pattern is quite similar to that of $30^\circ$,
while the long-range quasi-periodic nature is completely lost and round to a periodic pattern.
Here we confirmed that the quasi-band structures, DOS and IPR look almost the same as $30^\circ$,
but the tiny change of the wave bases and the coupling matrix elements in the 12-ring Hamiltonian
encodes the periodic / quasi-periodic transition.

The energy spectrum of the QC-TBG approximant can also be calculated by the original real-space tight-binding model
since it has a finite superlattice unit cell. We can show that the DOS and the wave function of $29.84^\circ$ 
calculated by the original tight-binding model are virtually the same as the result of the 12-ring effective Hamiltonian,
and this justifies the validity the effective approach.
In the appendix \ref{SI_quasicrystal_approximant}, we present the extensive study on the electronic structures of the quasi approximants
in all the angle region from $0^\circ$ to $30^\circ$.

\begin{figure}
	\begin{center}
		\leavevmode\includegraphics[width=1.\hsize]{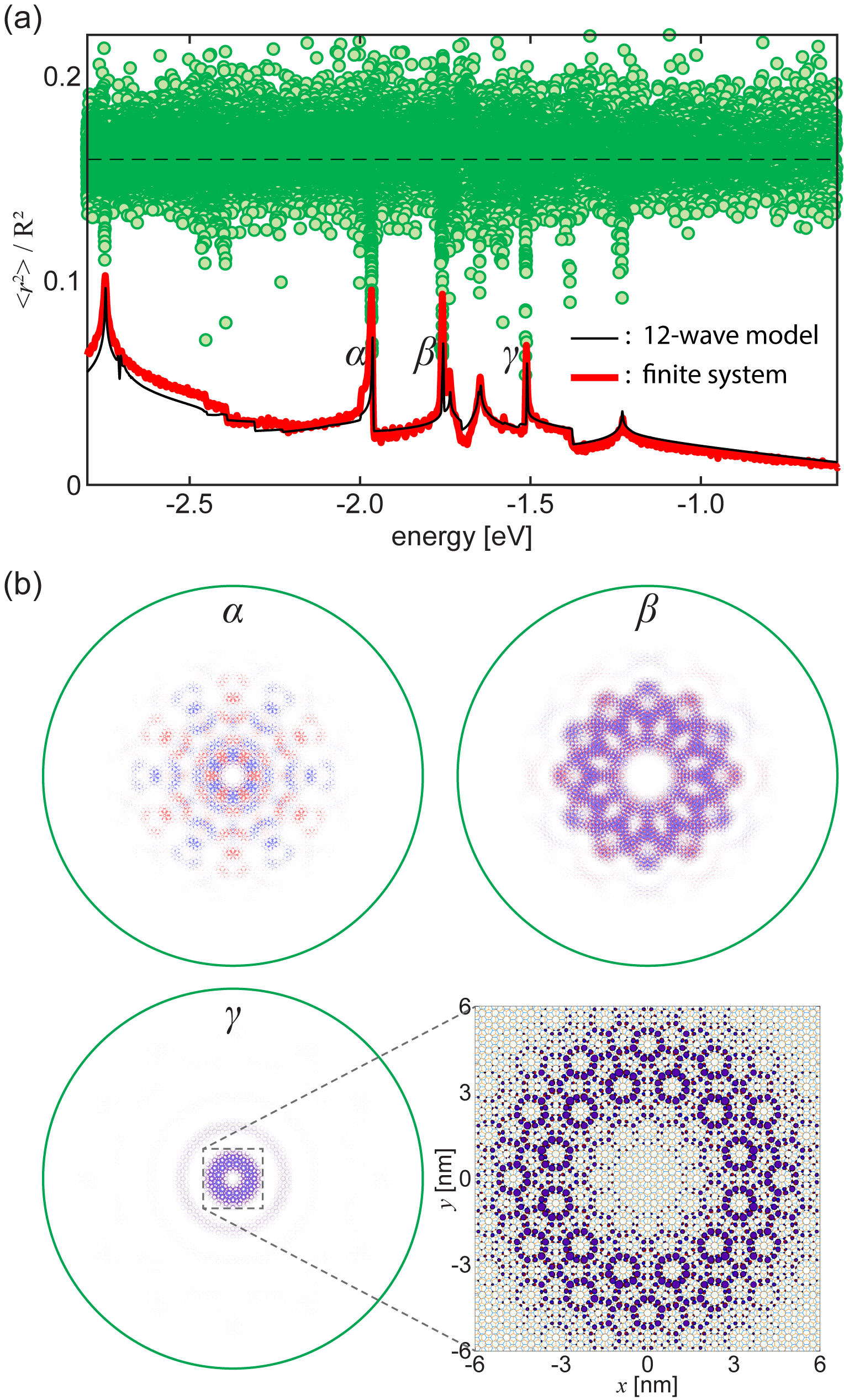}
	\end{center}
	\caption{
		(a) DOS (red line) and the generalized second momentum
		(green filled circles) of
		the two large finite flakes of graphene,
		with 371,532 atoms (radius of the flake $\sim$ $39.4\,\mathrm{nm}$),
		stacked at exactly $30^\circ$.
		Black line shows the DOS of the k-space model [12-wave model ($k_c < 18.8/a$)].
		(b) Plots similar to Fig.\,\ref{fig_12g_vb}(d)
		for each peak $\alpha$, $\beta$, $\gamma$.
		}
	\label{fig_flake_main}
\end{figure}

%%%
\subsection{Localization in finite-sized QC-TBG}
\label{sec_finite}

The emergence of quasicrystalline states in QC-TBG can also be confirmed by a finite-sized tight-binding lattice, while the computation is enormous.
Here we consider a tight-binding lattice composed of two large disks of graphene with radius $R = 39.4\,\mathrm{nm}$ 
stacked at exactly $30^\circ$, and calculate its electronic structures by diagonalizing the huge Hamiltonian matrix
with the total number of atoms 371,532.
As shown in Fig.\,\ref{fig_flake_main}(a),
the DOS of the finite flakes (thick red line), which is obtained by broadening its discrete spectrum,
is consistent with the DOS of the 12-wave effective model (thin black line) calculated by
the effective Hamiltonian with a few wave bases [Fig.\,\ref{fig_12g_vb}(b)].
In Fig.\,\ref{fig_flake_main}(b), we also present the wave functions at three energies, $\alpha$, $\beta$ and $\gamma$,
which correspond to the band edges of the quasi-band structures in the effective Hamiltonian (Fig.\,\ref{fig_12g_vb}).

The magnified plot of $\gamma$ is presented in
the inset of Fig.\,\ref{fig_flake_main}(b)
showing the characteristic pattern of 12-wave approximation.
Interestingly, however, it is overlapped with an envelope function decaying in the radial direction.
Such the localized feature is never seen in single layer graphene
and it is the characteristics of the resonant states of QC-TBG.
As a measure of the concentration to the center, we calculate the second momentum $\langle  r^2  \rangle / R^2$ for each eigenstate
and plot it as green circles in Fig.\,\ref{fig_flake_main}(a).
For a uniform state (i.e., the wave amplitude is constant throughout the system), 
$\langle  r^2  \rangle / R^2$ approaches $1/(2\pi)$, which is indicated by the dashed line.
We can actually see that $\langle  r^2  \rangle / R^2$ lies around this line for most of the states,
while it becomes exceptionally small at the energies of the quasi-band edges argued in the previous section.
In terms of the quasi-band structures,
these localized states actually correspond to the integral of the quasi-band states over the nearly flat region,
and the length scale of the envelope function is related to the size of the flat area in the momentum space.
For the state at $\gamma$ 
($m=0$ state), 
for instance, the radius of the flat area is roughly given by $\delta k \sim 0.2/a$, 
and the corresponding real-space scale $r = 2\pi/k \sim 7.7\,\mathrm{nm}$ matches the characteristic decaying and oscillating scale of the envelope function.

\section{Conclusions}
\label{sec_concl}

We revealed that the quasicrystalline nature emerges in the electronic properties of QC-TBG, 
or the twisted bilayer graphene stacked at 30$^\circ$.
We developed a concise model Hamiltonian for this unique system,
and demonstrated that the electronic structure is well described by the quasi-band picture despite of the lack of periodicity.
The quasi-band states of the QC-TBG are characterized by the 12-fold resonant states of relativistic Dirac fermions,
where the wave functions exhibit the spatial pattern fully respecting the dodecagonal quasicrystal tiling.
Such a non-uniform distribution of electron may be observed by microscopy imaging techniques.
The emergence of quasicrystalline states was attested by comparing the QC-TBG and a periodic approximant near $30^\circ$,
and it was demonstrated that even a slight deviation from the QC configuration destroys the long-range quasicrystalline nature.
Finally, we studied the electronic states of QC-TBG using the finite-size tight-binding model,
where the 12-fold resonant states appear as spatially-localized states in a finite-size geometry.

While we considered the QC-TBG as a model example in this paper,
the theoretical method based on the $k$-space tight-binding approach introduced here
is applicable to any kind of
extrinsic quasicrystals composed two-dimensional materials overlaid in incommensurate configurations, including heterostructures of two-dimensional materials having difference lattice symmetries (e.g., rectangle and hexagon).
%incommensurately stacked atomic layers,
%such as the stack of layers with two difference symmetries (e.g., rectangle and hexagon),
%extrinsic quasicrystals,
%and even the TBGs with any arbitrary $\theta$ without relying on the moir\'{e} periodicity.
%The detailed studies on exotic electronic natures in a broad class of extrinsic quasicrystals are left for future research.

Extrinsic quasicrystals also provide a unique opportunity to tune the quasicrystal bands
by controlling the interlayer interaction strength $U$.
As $U$ is an exponential function of the interlayer spacing $d$, \cite{DWNT}
we can either increase $U$ by applying pressure,
or decrease it through intercalation of ions or
addition of barrier atomic layers \cite{Chittari_2018}.
%Here we demonstrated that 
%the band flatness as well as the energy scale of the quasicrystal bands
%rapidly grow as $U$ increases.
When $U$ becomes comparable to the width of the energy bands,
we expect a transition from the weakly coupled regime
to the strongly coupled regime where the quasicrystalline nature is even more pronounced.
%\textcolor{red}{[[(??) Besides $U$, the control of the interlayer potential asymmetry
%or the sublattice symmetry will also tune the resonant interaction
%and enable the quasicrystal band structure engineering.]]}
%The growth of the band flatness will significantly reduces the conductivity,
%and the electron-electron interaction in such flat bands
%may serve as the source of many interesting phenomena
%such as the enhancement of electron-phonon coupling as well as superconductivity.
The detailed studies on exotic electronic natures in a broad class of extrinsic quasicrystals,
such as the electronic transport, optical properties, the quantum Hall effect,
and also the effects of $U$ modulation to these phenomena, are left for future research.

\section*{Acknowledgments}

%\item[Acknowledgments]
We thank L. A. Wray and A. Kent for fruitful discussions.
P.M. was supported by NYU Shanghai (Start-Up Funds), NYU-ECNU Institute of Physics at NYU Shanghai,
New York University Global Seed Grants for Collaborative Research.
This research was carried out on the High Performance Computing resources at NYU Shanghai and CAC of KIAS.
M.K. was supported by JSPS KAKENHI Grant Numbers JP25107005, JP15K21722, JP17K05496. Y.-W.S was supported by NRF of Korea (Grant No. 2017R1A5A1014862, SRC program: vdWMRC center).

\appendix

\section{Validity of the momentum space cut-off}
\label{SI_validity}
%\subsection{Localization in momentum space}
%\label{SI_localization_in_momentum_space}

\begin{figure*}
	\begin{center}
		\leavevmode\includegraphics[width=0.7\hsize]{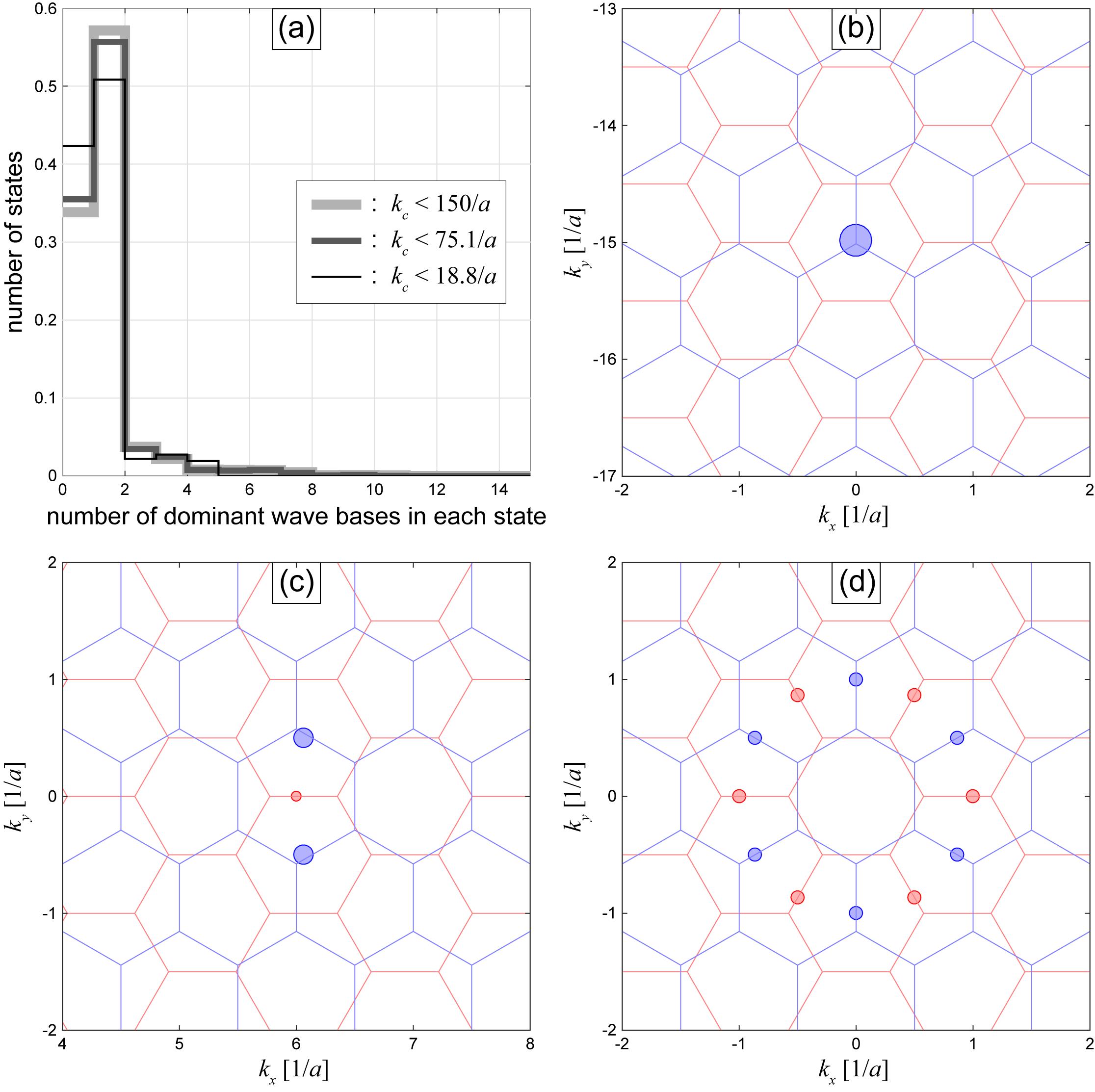}
	\end{center}
	\caption{(a) Histogram of the number of dominant wave bases component ${|\Vec{k},X\rangle}$ and ${|\tilde{\Vec{k}},X\rangle}$
%		over the entire number of bases composing each state.
		that make up each state.
		Thick-lightgray, middle-darkgray, thin-black lines show the histogram for all the states calculated with the wave bases within $k_c$ of $150/a$, $75.1/a$, $18.8/a$, respectively,
		where the total number of wave bases are $11630$, $2918$, $182$, respectively.		
		The histogram is normalized by the total number of states, which is two times the number of wave bases due to the sublattices.
		(b-d) The dominant component wave bases of three example states in the $k$-space.
		The radius of each shaded red (blue) circles is proportional to the amplitude of each wave basis in layer 1 (2).
		(b) The wave component of a nearly decoupled, monolayer-like state.
		(c) The state which originates from the 2-wave mixing. And,
%		showing that two monolayer states in the same layer interact with each other with a weak assist of the monolayer state in another layer,
		(d) The 12-wave resonance state.
	}
	\label{fig_localization_in_momentum_space}
\end{figure*}

In this section, we argue about the validity of introducing the momentum space cut-off 
in calculating the quasi band structure.
As we mentioned in Sec.\ \ref{sec_dual_tb}, the wave functions of the QC-TBG are localized in the $k$-space
in a similar manner to the Aubry-Andr\'{e} model in one-dimension \cite{aubry1980analyticity},
because the hopping term in the $k$-space is much smaller than the potential landscape (the band energy).
%\begin{align}
%\mathcal{H} &=J \sum_{m} (|w_m\rangle \langle w_{m+1}| + |w_{m+1}\rangle \langle w_m|) \nonumber\\
%&+ \Delta \sum_{m} \cos(2\pi\beta m+\phi)|w_m\rangle \langle w_m|
%\end{align}
%with the interaction strength $J$ between the neighboring states $|w_m\rangle$
%and the width of the potential $\Delta$,
%shows a sharp transition from extended to localized states at $\Delta/J = 2$.
Figures \ref{fig_localization_in_momentum_space}(b)-(d) show
some examples of the $k$-space amplitude map.
The panel (b) shows a nearly decoupled state which is dominated by only a single state of monolayer graphene,
and (c) is a state originating from the 2-wave mixing,
where a pair of monolayer's states on layer 2 are coupled though the mediation of a
middle state on layer 1. The panel (d) is the 12-wave resonant state.
Any eigenstates other than those examples
are also localized within just a few reciprocal lattice constants in $k$-space.
When we increase the number of total wave bases components
(${|\Vec{k},X\rangle}$ and ${|\tilde{\Vec{k}},\tilde{X}\rangle}$) by increasing $k_c$, 
each eigenstate hardly changes as long as $k_c$ is greater than the typical localization length.
In Fig.\,\ref{fig_localization_in_momentum_space}(a),
we show the histogram of the number of dominant wave components in the eigenstates
at a particular $\Vec{k}_0$,
calculated in the basis sets within $k_c$ of $150/a$, $75.1/a$ and $18.8/a$, respectively,
where the total number of wave bases are $11630$, $2918$ and $182$, respectively.	
We actually see that each of eigenstates is composed only a few (mostly less than 10) bases.
We note that, in large $k_c$,  we often see a resonance between different localized states 
which are very distant in $k$-space. This does not much affect the calculation of the physical quantity
because the overlap of the different localized wave functions are exponentially small.

%\subsection{Effects of shifting $\Vec{k}_0$}
%\label{SI_effects_of_shifting_k0}

We have infinitely many localized states far away from the first Brillouin zone,
so one might think that it is necessary to take an infinite $k_c$ to properly include all the states.
Note that, however, these localized states can be moved into the vicinity of the first Brillouin zone
by shifting $\Vec{k}_0$ with a proper amount, as we show in the following.
Thus, instead of using a large $k_c$ requiring a large computational cost,
we can obtain the full spectrum of the system by calculating the electronic structures
as a function of $\Vec{k}_0$ with a moderate $k_c$.

%The validity of discarding the wave points outside the cut-off circle with a radius $k_c$
%	in Sec.\ \ref{sec_dual_tb}
%	is based on the localization of the wave functions in the $k$-space, as well as
%	the fact that
%	the set of interacting wave points outside $k_c$ can be moved inside the first Brillouin zone
%	by shifting $\Vec{k}_0$.
	Let us consider two states $|\Vec{k}_1,X\rangle$ and $|\tilde{\Vec{k}}_1,\tilde{X}\rangle$ with
	\begin{align}
	\Vec{k}_1 &= \Vec{k}_0 + \tilde{\Vec{G}}_1 \quad (\tilde{\Vec{G}}_1 \in \tilde{\Vec{G}}), \nonumber\\
	\tilde{\Vec{k}}_1 &= \Vec{k}_0 + \Vec{G}_1 \quad (\Vec{G}_1 \in \Vec{G}),
	\label{eq_si_c_01}
	\end{align}
	for a given $\Vec{k}_0$.
	Suppose $\Vec{k}_1$ and $\tilde{\Vec{k}}_1$ are outside the cut-off circle,
	i.e., $|\Vec{k}_1| > k_c$ and $|\tilde{\Vec{k}}_1| > k_c$,
	but they strongly interact with each other, i.e.,
	\begin{equation}
	|\Vec{q} (= \Vec{k}_0 + \Vec{G}_1 + \tilde{\Vec{G}}_1)| \le \mathcal{O}(|\Vec{a}_i^*|).
	\label{eq_si_c_02}
	\end{equation}

	Now, for any such $\Vec{k}_1$, we can always find $\Vec{G}_2$ ($\Vec{G}_2 \in \Vec{G}$)
	which makes $\Vec{k}_1$ move to the point $\Vec{k}_2 \equiv \Vec{k}_1 - \Vec{G}_2$ in the first Brillouin zone, i.e.,
	\begin{equation}
	|\Vec{k}_2| \le \mathcal{O}(|\Vec{a}_i^*|).
	\label{eq_si_c_03}
	\end{equation}
	And suppose $\tilde{\Vec{k}}_2$, defined as
	\begin{equation}
	\tilde{\Vec{k}}_2 \equiv \Vec{k}_0 + \Vec{G}_1 + \tilde{\Vec{G}}_1.
	\label{eq_si_c_04}
	\end{equation}
	Then, by shifting $\Vec{k}_0$ to a new point $\Vec{k}_0'$ defined as
	\begin{equation}
	\Vec{k}_0' \equiv \Vec{k}_2,
	\label{eq_si_c_05}
	\end{equation}
	we can see that
	\begin{align}
	\Vec{k}_2 &= \Vec{k}_0' + \Vec{0} \quad (\Vec{0} \in \tilde{\Vec{G}}), \nonumber\\
	\tilde{\Vec{k}}_2 &= \Vec{k}_0' + \Vec{G}_1 + \Vec{G}_2 \quad (\Vec{G}_1 + \Vec{G}_2 \in \Vec{G}),
	\label{eq_si_c_06}
	\end{align}
	are the member of the subspace spanned from $\Vec{k}_0'$.
	And by considering that $\tilde{\Vec{k}}_2 = \Vec{q}$,
	and from Eqs.\,\ref{eq_si_c_02} and \ref{eq_si_c_03},
	we can show that these two points are within the cut-off circle.
	Since
	\begin{align}
	\Vec{k}_2 (= \Vec{k}_0 + \tilde{\Vec{G}}_1 - \Vec{G}_2) &= \Vec{k}_1 (=\Vec{k}_0+\tilde{\Vec{G}}_1) \quad \pmod{\Vec{G}}, \nonumber\\
	\tilde{\Vec{k}}_2 (= \Vec{k}_0 + \Vec{G}_1 + \tilde{\Vec{G}}_1)  &= \tilde{\Vec{k}}_1 (=\Vec{k}_0+\Vec{G}_1) \quad \pmod{\tilde{\Vec{G}}},
	\end{align}
	$|\Vec{k}_2,X\rangle$ and $|\tilde{\Vec{k}}_2,\tilde{X}\rangle$ represent
	the Bloch states same to $|\Vec{k}_1,X\rangle$ and $|\tilde{\Vec{k}}_1,\tilde{X}\rangle$, respectively,
	interacting with the same interaction strength $t(\Vec{q})$, since
	\begin{equation}
	\Vec{q}' \equiv \Vec{k}_2 - (\Vec{k}_0' - \tilde{\Vec{k}}_2) = \Vec{q}.
	\end{equation}
	Thus, by shifting $\Vec{k}_0$ to $\Vec{k}_0'$,
	the points discarded outside $k_c$ with $\Vec{k}_0$ are properly considered.
	And by calculating the electronic structures for every $\Vec{k}_0$ in the first Brillouin zone,
	we can get every possible interaction pairs in this system.

%\subsection{Replicas of identical quasi-bands}
%\label{SI_replicas_of_identical_quasibands}
%\section{Convergence of density of states with respect to the cut-off radius}

\begin{figure}
	\begin{center}
		\leavevmode\includegraphics[width=0.9\hsize]{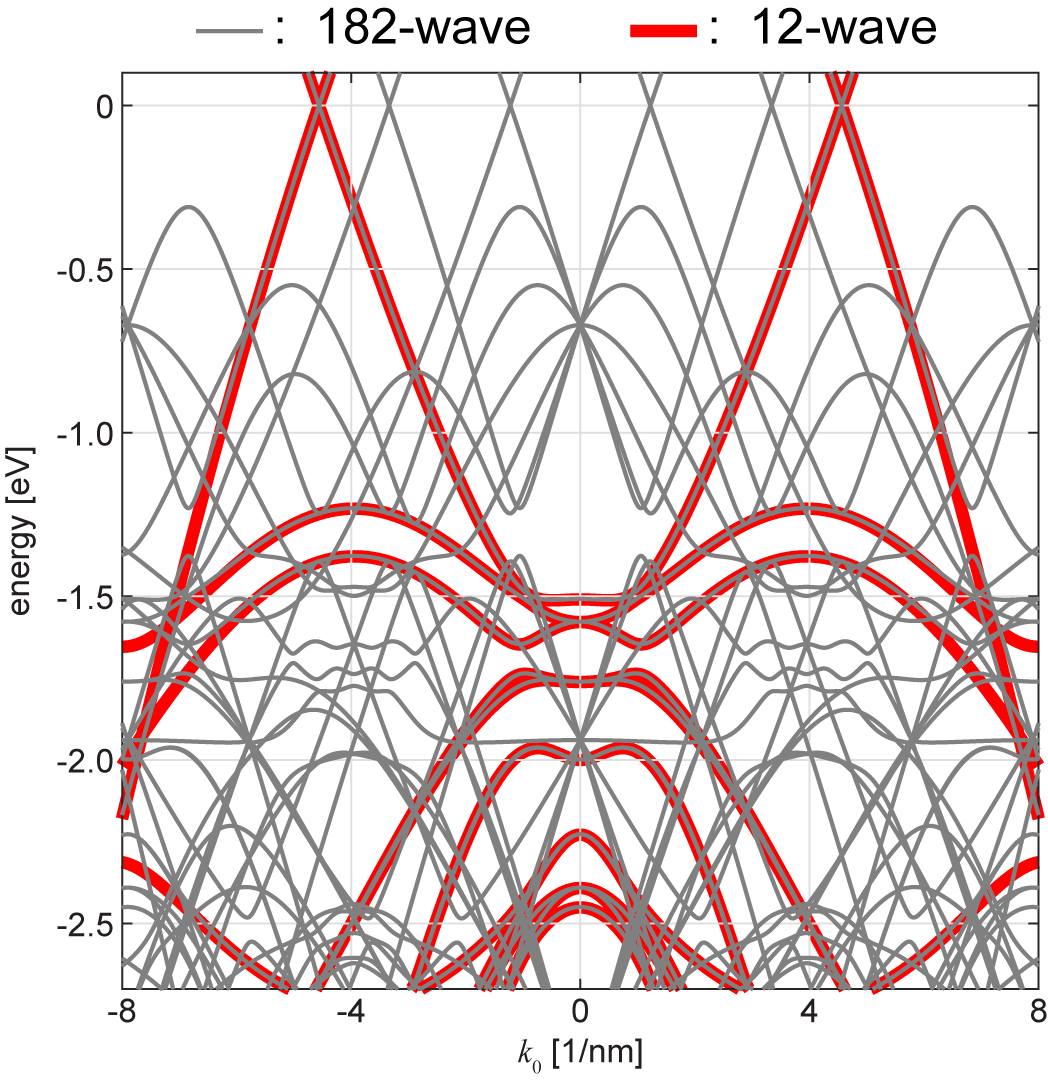}
	\end{center}
	\caption{Quasi-band structure of QC-TBG
			calculated by 12-wave model (thin red lines) and 182-wave model (thick gray lines).
	}
	\label{fig_effects_of_kc_1D}
\end{figure}

Figure \ref{fig_effects_of_kc_1D} shows the quasicrystal bands of QC-TBG
calculated with 12-wave model ($k_c < 3.76/a$ with $\Vec{k} = \Vec{0}$ removed)
and 182-wave model ($k_c < 18.8/a$).
We can see that the band structure of 182-wave model fully includes the spectrum of 12-wave model,
while also contains many other band lines.
Actually, these extra lines are just replicas of the identical quasicrystal bands of 12-wave model
with different origins (Sec.\ \ref{sec_dual_tb}).
In other words, the physical properties can be well described by calculating the quasi-band structure with
a relatively short $k_c$.

\begin{figure}
	\begin{center}
		\leavevmode\includegraphics[width=0.9\hsize]{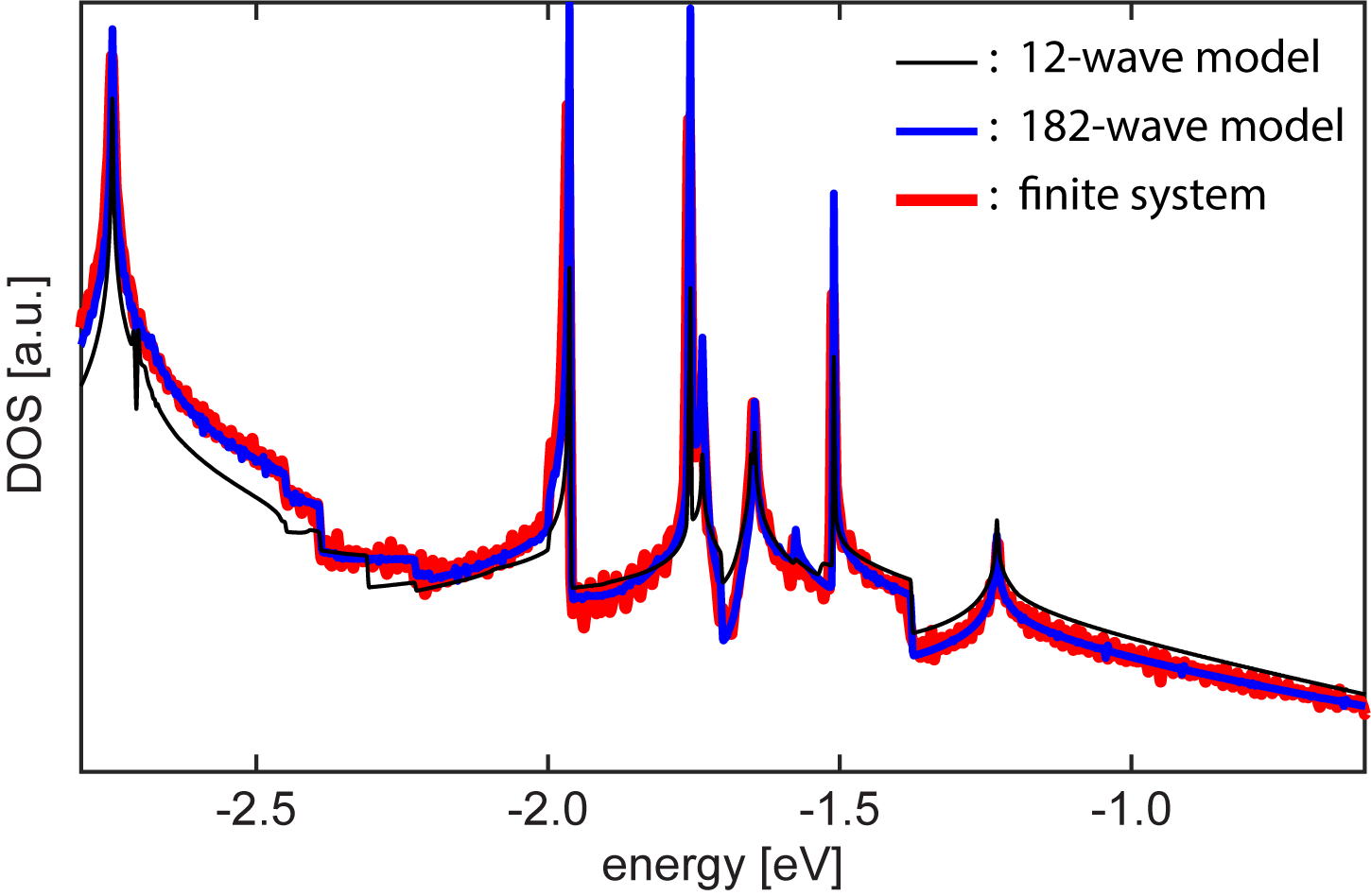}
	\end{center}
	\caption{DOS of QC-TBG calculated by 12-wave bases (thin black line) and 182-wave bases (middle blue line).
			The thick red line shows the DOS of finite-sized QC-TBG.
	}
	\label{fig_SI_DOS}
\end{figure}

%Almost every states of QC-TBG are localized to less than 10 wave bases in the $k$-space (appendix \ref{SI_localization_in_momentum_space}).
%The states right on the boundary of $k_c$ might show some artifacts, since the interaction between the wave bases composing the states are cut by $k_c$.
%Note that, however, such artifact can be averaged out by taking sufficiently many $k_0$, since shifting $k_0$ will move such states away from $k_c$
%(appendix \ref{SI_effects_of_shifting_k0}).
%Thus, we need to choose $k_c$ sufficiently larger than the localization length, but it needs not be too large.
The minimum 12-wave models well reproduces the band structure near the 12-wave resonant states,
while there are some small errors in the other energies.
In Fig.\,\ref{fig_SI_DOS},
we see that the 182-wave model almost perfectly overlaps with the DOS of very large finite flakes,
while 12-wave model slightly under-/overestimates the density of states far from the resonant-state peaks.
We confirmed that further increase of $k_c$ does not change the DOS profile.

\section{Quasicrystal approximants}
\label{SI_quasicrystal_approximant}

In Sec.\ \ref{sec_12-fold}, we compared the quasicrystalline TBG stacked at $30^\circ$ (QC-TBG)
and its periodic approximant at $29.84^\circ$.
Actually there exist infintely many periodic TBGs 
in any finite region in $\theta$, just like rational numbers in the real number axis.
As we will see the following, the peak structure in the density of states 
changes almost continuously in rotating the twist angle $\theta$,
and tracing its evolution is useful to get insights on the connection between QC-TBG and the low-angle moir\'{e} TBGs,
although the computation requires enormous number of atomic bases ($10^3-10^5$ atoms).

\begin{figure*}
	\begin{center}
		\leavevmode\includegraphics[width=0.7\hsize]{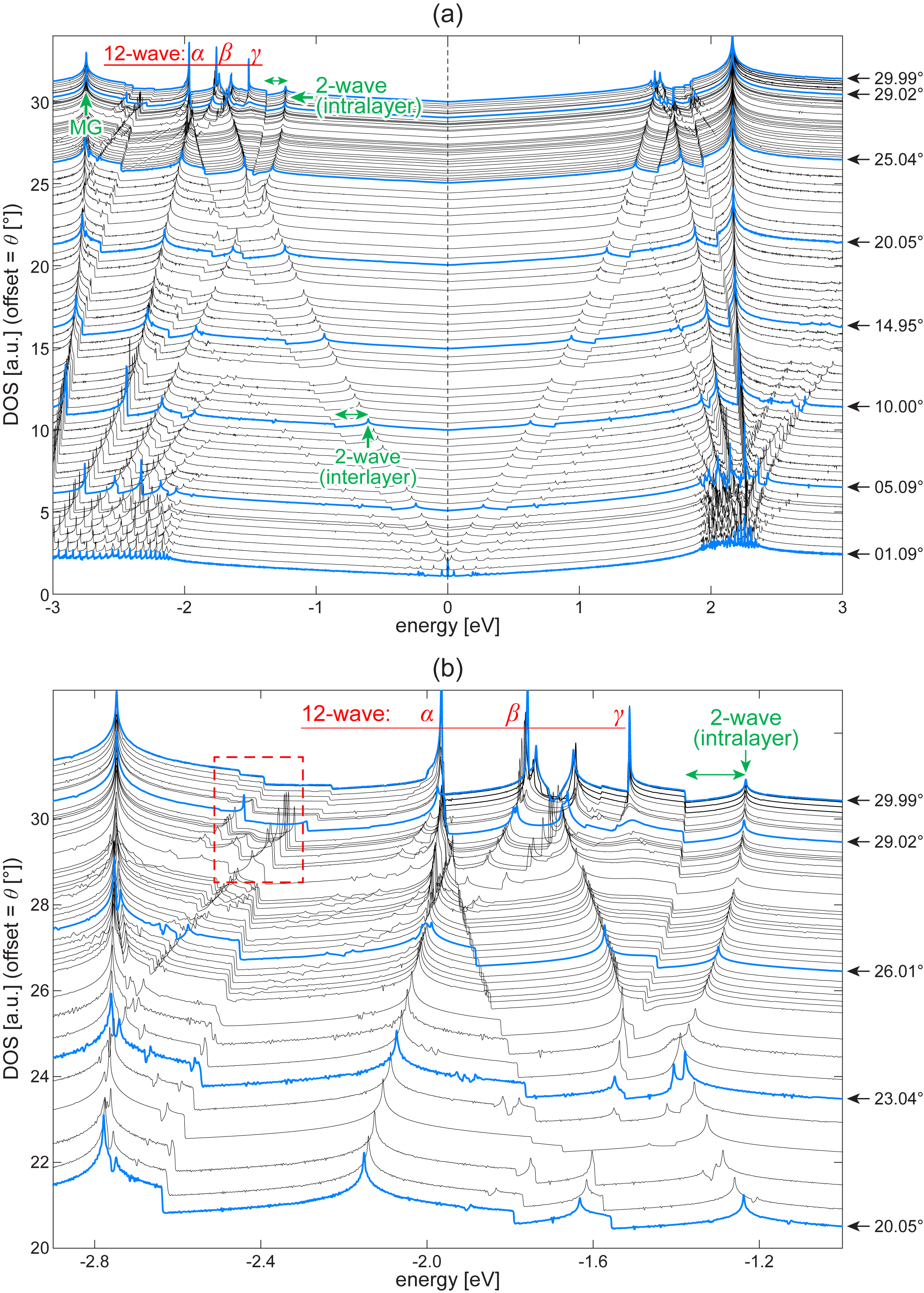}
	\end{center}
	\caption{(a) DOS of TBGs with various rotation angles $0^\circ < \theta < 30^\circ$ in a wide range of energy.
		Each line is offset along the vertical direction by $\theta$.
		Peaks marked with green arrows correspond to the van Hove singularity of monolayer graphene,
		and two-wave mixing (interlayer and intralayer) of TBGs, respectively.
		Peaks marked as $\alpha$, $\beta$, $\gamma$ correspond to the singularities
		associated with the resonant states of QC-TBG ($\theta = 30^\circ$).
		(b) Magnified plot of (a) near the energy ranges of the critical states of QC-TBG.
	}
	\label{fig_approximant}
\end{figure*}

Here we calculated the electronic structures of periodic TBGs
with various $\theta$ using a tight-binding model.
Figure \ref{fig_approximant}(a) shows the evolution of the DOS in a wide range of energy,
and Fig.\,\ref{fig_approximant}(b) is the magnified plot near the critical states of QC-TBG.
The peaks marked with ``MG" correspond to the van Hove singularity of monolayer graphene.
Similarly, ``2-wave-interlayer" represents the singularity originates from the two-wave mixing
between the states in different layers  \cite{moon_optics},
and ``2-wave-intralayer" is the mixing between the states in the same layer. \cite{DWNT,Yao2018} 
We show the width of the band opening (pseudogap) from 2-wave interlayer/intralayer mixing
by the green horizontal arrows.

The sharp peaks marked as $\alpha$, $\beta$, $\gamma$ correspond to the singularities 
coming from the 12-wave mixing of QC-TBG, which were described as the nearly flat bands in the quasi-band picture
in Sec.\ \ref{sec_12-fold}.
We can see that the singular peaks rapidly grow as $\theta$ approaches $30^\circ$.
The DOS of the TBG with $\theta = 29.99^\circ$, the periodic TBG closest to 30$^\circ$ in this 
calculation, is consistent with that of QC-TBG [Fig.\,\ref{fig_12g_vb}(a)].
It should be noted that, however,
the wave functions of the approximants do not obey the quasicrystalline long-range structure
with 12-fold rotational symmetry,  as argued in Fig.\,\ref{fig_ldos_symmetry_approximant_and_ecm}(b).
We also see that the peak-and-dip structure in the valence band is much wider than
in the conduction band, and it is consistent with the analytic argument in the 12-wave ring model 
[Eq.\,(\ref{eq_Em})].

The peaks enclosed by the red box in Fig.\,\ref{fig_approximant}(b)
are associated with the resonant states other than $\alpha$, $\beta$, $\gamma$
[i.e., the solutions of Eq.\,(\ref{eq_H12}) other than $\alpha$, $\beta$, $\gamma$].
In the 12-wave model, we can also show that corresponding states have flat dispersion in quasi-band structure
at $\theta \sim 28^\circ$, and exhibit singularities in DOS.
As $\theta$ approaches $30^\circ$, however, the quasi-band becomes dispersive [Fig.\,\ref{fig_12g_vb}(b)]
and the DOS singularities disappear.

\bibliography{wqc}
\end{document}